\documentclass[11pt]{article}

\newsavebox{\foobox}
\newcommand{\slantbox}[2][0]{\mbox{%
        \sbox{\foobox}{#2}%
        \hskip\wd\foobox
        \pdfsave
        \pdfsetmatrix{1 0 #1 1}%
        \llap{\usebox{\foobox}}%
        \pdfrestore
}}
\newcommand\unslant[2][-.25]{\slantbox[#1]{$#2$}}

\newcommand{\mpi}{\text{\unslant[-.18]\pi}}
\newcommand{\mdelta}{\text{\unslant[-.18]\delta}}

\renewcommand\textfraction{.05}

\usepackage[left=2cm, right=2cm, top=2.5cm, bottom=2.5cm]{geometry}
\usepackage[x11names]{xcolor}
\usepackage{fancyhdr, amssymb, cancel, amsmath, graphicx, pgfplots, tikz}
\usepackage{isomath}

\usetikzlibrary{shadows}

\newcommand{\stylecolor}{violet}

\usepackage[labelfont={bf,sf, color=\stylecolor}, margin={1.5cm,0cm}]{caption}

\usepackage[colorlinks=true, urlcolor=\stylecolor!70!white, linkcolor=\stylecolor, citecolor=\stylecolor!70!white, hyperindex=true, linktocpage=true]{hyperref}

\usepackage[explicit]{titlesec}

\newcommand*\sectionlabel{}
\titleformat{\section}
  {\gdef\sectionlabel{}
   \Large\bfseries\scshape}
  {\gdef\sectionlabel{\thesection }}{0pt}
  {\begin{tikzpicture}[remember picture,overlay]
       \end{tikzpicture}
  }
\titlespacing*{\section}{0pt}{0pt}{0pt}

\newcommand*\subsectionlabel{}
\titleformat{\subsection}
  {\gdef\subsectionlabel{}
   \large\bfseries\scshape}
  {\gdef\subsectionlabel{\thesubsection  }}{0pt}
  {\begin{tikzpicture}[remember picture,overlay]
    	\draw (-0.15, 0.02) node[right] {\color{\stylecolor} \textsf{\subsectionlabel}};
	\draw (1.25, 0) node[right] {\color{\stylecolor} \textsf{#1}};
	\fill[color=\stylecolor] (1,-0.25) rectangle (1.1, 0.25);
       \end{tikzpicture}
  }
\titlespacing*{\subsection}{0pt}{10pt}{10pt}

\newcommand*\subsubsectionlabel{}
\titleformat{\subsubsection}
  {\gdef\subsubsectionlabel{}
   \bfseries\scshape}
  {\gdef\subsubsectionlabel{\thesubsubsection.\ \  }}{0pt}
  {\begin{tikzpicture}[remember picture,overlay]
    	\draw (-0.15, 0) node[right] {\color{\stylecolor} \textsf{\subsubsectionlabel#1}};
       \end{tikzpicture}
  }
\titlespacing*{\subsubsection}{0pt}{7pt}{7pt}

\pgfplotsset{every axis legend/.append style={at={(1.02,1)},anchor=north west}}

\begin{document}

\allowdisplaybreaks

\renewcommand{\labelitemi}{\color{\stylecolor} $\blacktriangleright$}

\setcounter{totalnumber}{5}
\renewcommand\textfraction{.1}

\pagestyle{fancy}
\renewcommand{\headrulewidth}{0pt}
\fancyhead{}

\fancyfoot{}
\fancyfoot[C] {\textsf{\textbf{\thepage}}}

\begin{equation*}
\begin{tikzpicture}
\draw (\textwidth, 0) node[text width = \textwidth, right] {\color{white} easter egg};
\end{tikzpicture}
\end{equation*}

\begin{equation*}
\begin{tikzpicture}
\draw (0.5\textwidth, -3) node[text width = \textwidth] {\huge  \textsf{\textbf{Charge diffusion and the butterfly effect in striped \\ \vspace{0.07in} holographic matter}} };
\end{tikzpicture}
\end{equation*}
\begin{equation*}
\begin{tikzpicture}
\draw (0.5\textwidth, 0.1) node[text width=\textwidth] {\large \color{black} \textsf{Andrew Lucas}$^{\color{\stylecolor} \mathsf{a,b}}$ \textsf{and Julia Steinberg}$^{\color{\stylecolor} \mathsf{a}}$};
\draw (0.5\textwidth, -0.5) node[text width=\textwidth] { $^{\color{\stylecolor} \mathsf{a}}$  \small\textsf{Department of Physics, Harvard University, Cambridge, MA 02138, USA}};
\draw (0.5\textwidth, -1) node[text width=\textwidth] { $^{\color{\stylecolor} \mathsf{b}}$  \small\textsf{Department of Physics, Stanford University, Stanford, CA 94305 USA}};
\end{tikzpicture}
\end{equation*}
\begin{equation*}
\begin{tikzpicture}
\draw (0, -13.1) node[right, text width=0.5\paperwidth] {\texttt{ajlucas@stanford.edu} \\ \texttt{jsteinberg@g.harvard.edu}  };
\draw (\textwidth, -13.1) node[left] {\textsf{\today}};
\end{tikzpicture}
\end{equation*}
\begin{equation*}
\begin{tikzpicture}
\draw[very thick, color=\stylecolor] (0.0\textwidth, -5.75) -- (0.99\textwidth, -5.75);
\draw (0.12\textwidth, -6.25) node[left] {\color{\stylecolor}  \textsf{\textbf{Abstract:}}};
\draw (0.53\textwidth, -6) node[below, text width=0.8\textwidth, text justified] {\small  Recently, it has been proposed that the butterfly velocity -- a speed at which quantum information propagates -- may provide a fundamental bound on diffusion constants in dirty incoherent metals.   We analytically compute the charge diffusion constant and the butterfly velocity in charge-neutral holographic matter with long wavelength ``hydrodynamic" disorder in a single spatial direction.   In this limit, we find that the butterfly velocity does not set a sharp lower bound for the charge diffusion constant.};
\end{tikzpicture}
\end{equation*}

\tableofcontents

\titleformat{\section}
  {\gdef\sectionlabel{}
   \Large\bfseries\scshape}
  {\gdef\sectionlabel{\thesection }}{0pt}
  {\begin{tikzpicture}[remember picture,overlay]
	\draw (1, 0) node[right] {\color{\stylecolor} \textsf{#1}};
	\fill[color=\stylecolor] (0,-0.35) rectangle (0.7, 0.35);
	\draw (0.35, 0) node {\color{white} \textsf{\sectionlabel}};
       \end{tikzpicture}
  }
\titlespacing*{\section}{0pt}{15pt}{15pt}

\begin{equation*}
\begin{tikzpicture}
\draw[very thick, color=\stylecolor] (0.0\textwidth, -5.75) -- (0.99\textwidth, -5.75);
\end{tikzpicture}
\end{equation*}

\section{Introduction}
One of the simplest properties of a metal to measure is its electrical conductivity $\sigma$.   In an ordinary metal like iron or alumnium, Fermi liquid theory predicts that $\sigma \sim T^{-2}$, and indeed at low temperatures this scaling can be confirmed for most metals \cite{ashcroft}.   However, there is a long-standing experimental puzzle in which certain ``strange metals" instead have $\sigma \sim T^{-1}$ \cite{mackenzie2013}.   Unlike ordinary Fermi liquids, these strange metals are believed to be strongly correlated, making it difficult to find controllable theoretical models for them \cite{taillefer, keimer}.   

A few years ago, it was noted that one can use the classic Drude formula for quasiparticle transport \cite{ashcroft} to define a relaxation time $\tau$: \begin{equation}
\tau \equiv \frac{m}{ne^2} \sigma,  \label{eq:drude}
\end{equation} 
with $n$ the carrier density and $m$ a quasiparticle effective mass (measured by quantum oscillations).   One finds that for many strange metals \cite{mackenzie2013}: \begin{equation}
\tau \sim \frac{\hbar}{k_{\mathrm{B}}T}.  \label{eq:hkT}
\end{equation}
However, it is hard to take such a result seriously as a quasiparticle scattering time, because this time scale has been argued for some time to be the fastest possible time scale characterizing the dynamics in any interacting quantum system \cite{sachdev96, sachdev98, sachdev}.   This has recently been revisited in \cite{stanford1503}.    However, \cite{hartnoll2014} pointed out that it would not be unreasonable to demand \begin{equation}
\tau \gtrsim \frac{\hbar}{k_{\mathrm{B}}T},  \label{eq:hkT2}
\end{equation}
because if we take $\tau$ seriously as some type of ``relaxation time", this time scale should not be faster than the ``interaction time scale" (\ref{eq:hkT}).   In the absence of quasiparticles, \cite{hartnoll2014} further noted that it is easier to postulate that diffusion constants of charge and energy are related to a relaxation time $\tau$:  \begin{equation}
D \gtrsim v^2 \tau \gtrsim v^2 \frac{\hbar}{k_{\mathrm{B}}T},  \label{eq:diffLB}
\end{equation} 
where $v$ is a velocity scale.   This formula is better than (\ref{eq:drude}) since it does not depend on the existence of quasiparticles to make sense.   If $v$ was independent of temperature, then since $D\chi =  \sigma$, with $\chi$ the charge susceptibility, \cite{hartnoll2014} proposed that $\sigma \sim T^{-1}$ in strange metals because their charge diffusion constant saturates the universal lower bound (\ref{eq:diffLB}).   The universality of such a bound is appealing,  but without a precise conjecture for how to compute $v$, such a bound is not useful.   For example, if $v$ and/or $\chi$ has temperature-dependence, then (\ref{eq:diffLB}) can be satisfied while $\sigma$ does not scale as $T^{-1}$.

Assuming a conjecture for what $v$ is, one way to check such a bound rigorously is to use gauge-gravity duality \cite{koenraadbook, rmp}.    This technique allows us to access the physics of certain strongly interacting finite temperature and density quantum systems by mapping their dynamics onto classical gravity.   Indeed, early results \cite{kovtun2004}  out of gauge-gravity duality suggested a universal viscosity bound relating the shear viscosity $\eta$ to the entropy density $s$: \begin{equation}
\eta \gtrsim \frac{\hbar s}{k_{\mathrm{B}}}.  \label{eq:viscbound}
\end{equation}
A precise coefficient was found for the simplest holographic models \cite{kovtun2004} and was conjectured to be universal.   Such a viscosity bound can be understood by the following chain of logic: \begin{equation}
\eta \sim \epsilon \tau \gtrsim  \frac{\hbar \epsilon}{k_{\mathrm{B}}T}  \sim \frac{\hbar s}{k_{\mathrm{B}}}.
\end{equation} with $\epsilon$ the energy density.   More recently, it has been shown that it is possible to parametrically violate these viscosity bounds \cite{yaida07, yaida08, anv1, anv2, anv3, mrv1, mrv2, mrv3, mrv4}.   Nonetheless, (\ref{eq:viscbound}) qualitatively holds for a large number of theories, and has not been meaningfully violated experimentally \cite{adams}.   Hence, there is hope that the original conjecture (\ref{eq:viscbound}) is still an excellent frame for thinking about the viscosity of strongly interacting quantum systems, even if it is not precise.

Given a sharp conjecture for what the velocity scale $v$ is in (\ref{eq:diffLB}) and the zoo of holographic models which are now routinely studied \cite{koenraadbook}, it is natural to use holographic approaches to test (\ref{eq:diffLB}).    For the remainder of the paper, we will set $c=\hbar=k_{\mathrm{B}}=1$ for simplicity;  such units can straightforwardly be restored with dimensional analysis.

\subsection{Charge Diffusion and the Butterfly Effect}
The recent papers \cite{blakeB1, blakeB2} have proposed that $v$ should be interpreted as a velocity called the butterfly velocity $v_{\textsc{b}}$.    Let us specialize to the study of charge diffusion in a locally charge neutral quantum field theory.   In this case, \cite{blakeB1} computed $D$ in a large set of spatially homogeneous low temperature scaling theories and, in many cases, found the simple relation: \begin{equation}
D =  \mathcal{C} \frac{v_{\textsc{b}}^2}{T},
\end{equation}
with $\mathcal{C}$ an O(1) constant which depends on certain low energy scaling dimensions of the theory,  but not on further details.    If $v=v_{\textsc{b}}$, then (\ref{eq:diffLB}) suggests we should find \begin{equation}
D \gtrsim \frac{v_{\textsc{b}}^2}{T}  \label{eq:diffLB2}
\end{equation}
in more generic situations.    

Before we discuss the extent to which (\ref{eq:diffLB2}) holds, let us introduce what the butterfly velocity actually is.   Let us consider a generic quantum field theory with operator $V$, localized around $x=0$ and $t=0$, and $W$, localized around $x=x_0$ and $t=t_0$.     If $t_0=0$, then causality implies that $V$ and $W$ must be uncorrelated.   At later times, this need not be the case.  In strongly interacting systems, one will generically find \cite{bhbutterfly, localized} \begin{equation}
\frac{\langle V(0,0)W(x_0,t_0)V(0,0)W(x_0,t_0)\rangle_T}{\langle V(0,0)V(0,0)\rangle_T\langle W(x_0,t_0)W(x_0,t_0)\rangle_T} = 1 -  \mathrm{e}^{\lambda(t_0-t_* - |x_0|/v_{\textsc{b}})},  \label{eq:sec1vb}
\end{equation}  
where $t_*$ is a time scale called the ``scrambling time", and $v_{\textsc{b}}$ is a velocity scale called the butterfly velocity.    When $x_0=0$,  this equation is reminiscent of chaos theory.   In a chaotic system, a small perturbation can grow exponentially large at late times, and so two nearly identical initial conditions can lead to dramatically different outcomes:  this is called the butterfly effect.   $\lambda$, identified as a Lyapunov exponent, tells us the rate at which the quantum system can become ``scrambled" and lose memory of its initial state.   Recently, \cite{stanford1503} has pointed out that under plausible physical assumptions, \begin{equation}
\lambda \le 2\mpi T.  \label{eq:lambda}
\end{equation}
This gives a precise meaning to (\ref{eq:hkT2}).   When $x_0\ne 0$, there is a further spatial delay in scrambling -- this is what is captured by the butterfly velocity.   One can crudely think of it as the speed at which quantum information can propagate through the system.   More precisely, \cite{lrbutterfly} has suggested identifying the butterfly velocity with an effective Lieb-Robinson velocity -- this latter velocity scale has been of great importance in quantum information theory for many years.   For our purposes, the precise quantum information theoretic interpretation of $v_{\textsc{b}}$ is not important -- as emphasized in \cite{blakeB1}, we focus on $v_{\textsc{b}}$ since it is a natural velocity scale to define in a strongly-coupled quantum system.

\subsection{Breaking Translation Symmetry}
We will test (\ref{eq:diffLB2}) in charge neutral theories without translation symmetry.  There are two reasons why such a test is important.   Firstly, to the extent one can even define viscosity in such systems, translation symmety breaking parametrically spoils the celebrated viscosity bound (\ref{eq:viscbound}) \cite{mrv1, mrv2, mrv3, mrv4}.   $D$ is well-defined in such theories as long as charge is conserved, but one might expect that translation symmetry breaking can spoil any bound.     Secondly, and more importantly, in a typical metal (which is at finite charge density) the only reason that the charge diffusion constant is not infinite is because impurities or umklapp processes break translation invariance.   When translation invariance is weak \cite{hkms, lucas1501, lucasMM}, $\sigma$ and $D$ are both parametrically large.   Hence, translation symmetry breaking must be a non-perturbatively strong effect in order for $D$ to have any chance of saturating a universal lower bound such as (\ref{eq:diffLB2}).    So although we will be studying charge neutral systems for computational simplicity, where $\sigma$ and $D$ are finite even in homogeneous systems, it is crucial that (\ref{eq:diffLB2}) be robust to translation symmetry breaking in order for it to be a sensible proposal.

There are some reasons to be optimistic about (\ref{eq:diffLB2}), at least in holographic models.   It has recently been shown that some classes of holographic models (in particular, the Einstein-Maxwell theory in four bulk spacetime dimensions) admit sharp conductivity bounds \cite{grozdanov, grozdanov2, ikeda}.\footnote{Such conductivity bounds must always be interpreted within the context of a specific holographic action:  namely, for any given \emph{theory}, the disorder profile cannot reduce the conductivity below a minimal value.   It is possible to, in addition, modify the boundary theory by modifying the bulk action, to decrease the conductivity further \cite{nobound1, nobound2}.   Since (\ref{eq:diffLB2}) contains theory dependent quantities on both sides of the inequality, there is hope that it may be more general.}   Such conductivity bounds can be saturated by the predictions of simpler models of disorder such as massive gravity \cite{blake1} or Q-lattice/linear axion models \cite{donos1, andrade, donos2, gouteraux}.    Hence, it may be the case that (\ref{eq:diffLB2}) is robust to translation symmetry breaking.   On the other hand, simple counterexamples to the conductivity bounds suggested \cite{xhge} by these simple models can readily be generated \cite{ikeda} even by simply changing the spacetime dimension.

In this paper, we compute $D$ and $v_{\textsc{b}}$ in inhomogeneous holographic matter coupled to a charge-neutral scalar operator $\mathcal{O}$.   If the field $H$ which linearly couples to $\mathcal{O}$ varies on long (hydrodynamic) spatial length scales $\xi$, we can construct the  dual geometry and compute $D$ and $v_{\textsc{b}}$ analytically.     Our computation demonstrates that there is no sharp lower bound for the charge diffusion constant, even in our simple class of models.  In fact, we find that in the same low temperature scaling regimes studied in \cite{blakeB1}:  \begin{equation}
D \le \mathcal{C} \frac{v_{\textsc{b}}^2}{2\mpi T},  \label{eq:diffLB3}
\end{equation}
with equality only holding in homogeneous backgrounds where $H$ is constant.     Hence, (\ref{eq:diffLB2}) does not generally hold with a precise constant prefactor.    While we do not present an explicit example where the temperature dependence of $D$ differs from the temperature dependence of $v_{\textsc{b}}^2/T$ at low temperature,  we also cannot rule this possibility out in exotic holographic matter.


The outline of this paper is as follows.   In Section \ref{sec2}, we review the fluid-gravity correspondence and describe the construction of the striped black holes we use to compute $v_{\textsc{b}}$ and $D$.     Sections \ref{sec3} and \ref{sec4} compute $D$ and $v_{\textsc{b}}$ respectively.  We find that both $D$ and $v_{\textsc{b}}$ admit, at leading order in $\xi^{-1}$, a simple hydrodynamic interpretation.   In the low-temperature limit of \cite{blakeB1}, we demonstrate (\ref{eq:diffLB3}) in Section \ref{sec5}.  Appendices contain further technical details.

\section{Striped Black Holes} \label{sec2}
In this paper, we will study asymptotically anti-de Sitter geometries in $d+2$ spacetime dimensions, dual to systems with a UV conformal fixed point: \begin{equation}
\mathrm{d}s^2(r\rightarrow 0) = g_{MN}\mathrm{d}x^M\mathrm{d}x^N = \frac{\mathrm{d}r^2 - \mathrm{d}t^2 + \mathrm{d}\mathbf{x}^2 }{r^2} \end{equation}
$r$ is the bulk radial coordinate, and $t$ and $\mathbf{x}$ represent time and space in the boundary field theory directions.    Without (yet) specifying the bulk matter content, we suppose that there is a scalar field $\Phi$, dual to a relevant operator of dimension $\Delta$ in the UV conformal field theory.    We consider states deformed by a spatially inhomogeneous source for the operator dual to $\Phi$.  The geometric interpretation of this is that the boundary conditions for $\Phi$ are: \begin{equation}
\Phi(x,r\rightarrow 0) =  \frac{1}{L^{d/2}} H(x) r^{d+1-\Delta} + \cdots.
\end{equation}
We wish to find a background geometry dual to a state of the theory at finite temperature $T$, assuming charge neutrality (at least for the charge whose diffusion constant we are computing).   For simplicity, we will suppose that the function $H(x)$ only depends on a single coordinate, which we will label as (italic) $x$.   To this charge neutral background, we then add a U(1) gauge field $A$, so the total action of the bulk theory is:  \begin{equation}
S = S_{\mathrm{matter}} - \int \mathrm{d}^{d+2}x \sqrt{-g} Z(\Phi) \frac{F^2}{4e^2},   \label{eq:sec2F2}
\end{equation}
with $F=\mathrm{d}A$.    The diffusion constant $D$ we compute will be the diffusion constant associated with the dual conserved charge.

In general,  the construction of such inhomogeneous geometries cannot be done analytically.   However, suppose that the function $H(x)$ varies on long wavelengths $\xi$:  i.e. \begin{equation}
\frac{\partial_x H}{H} \sim \frac{1}{\xi}
\end{equation}
and that the dimensionless parameter \begin{equation}
\frac{1}{\xi T} \rightarrow 0.  \label{eq:fglim}
\end{equation} In what follows, we will often discuss the perturbative expansion in $1/\xi$, and this will imply that $1/\xi T$ is the small dimensionless parameter governing the expansion.    In this case, we expect that the field theory -- deformed by a very long wavelength source -- reaches local thermal equilibrium as if it was in a homogeneous medium.   This ``hydrodynamic" disorder \cite{lucas, lucas3} is particularly elegant to treat holographically:  the fluid-gravity correspondence \cite{flugrav1, flugrav2} allows us to analytically construct inhomogeneous black hole backgrounds as a perturbative expansion in $1/\xi$: see e.g. \cite{chesler}.    This will allow us to analytically compute the charge diffusion constant $D$ and butterfly velocity $v_{\textsc{b}}$.   

At leading order in $1/\xi$, it is particularly simple to write down the inhomogeneous background geometry, with our specified boundary conditions for $\Phi$ and $g_{MN}$.   To do so, one must find a coordinate system which is regular at the horizon.   In such a regular coordinate system, we first construct the background geometry assuming that $H(x)= \text{constant}$.    One such coordinate system is Eddington-Finkelstein coordinates, where \begin{subequations}\begin{align}
\mathrm{d}s^2 &= -2\tilde a(r;H)\mathrm{d}v \mathrm{d}r  - a(r;H)\mathrm{d}v^2 + b(r;H) \mathrm{d}\mathbf{x}^2, \\
\Phi &= \varphi(r;H).
\end{align}\end{subequations}
 We have replaced here the standard time coordinate $t$, with a coordinate $v$ which is constant along in-falling light rays.\footnote{A more standard metric choice might have been $\mathrm{d}s^2 = g\mathrm{d}r^2 - f\mathrm{d}t^2 + h\mathrm{d}\mathbf{x}^2$.   These are often called Fefferman-Graham coordinates.   One can get from this coordinate system to Eddington-Finkelstein coordinates by defining $\mathrm{d}v = \mathrm{d}t - \sqrt{f/g} \mathrm{d}r $.   However, such metrics are not well-suited for the fluid-gravity correspondence, as we will explain shortly.}
The functions $a(r;H)$, $\tilde a(r;H)$ and $b(r;H)$ are functions of $H$, the boundary condition on the scalar $\Phi$, and also the temperature $T$ of the horizon.   Since black hole theorems imply that $T$ must be uniform across the horizon \cite{wald}, we only denote the explicit depenence on $H$, which is allowed to be spatially varying shortly.    Note that the coordinate $r$ may be reparameterized:  this implies that $\tilde a$ is not a true degree of freedom.  
For future reference, we note that the inverse metric has components \begin{equation}
g^{rr} = \frac{a}{\tilde a^2}, \;\;\; g^{rv} = \frac{1}{\tilde a}, \;\;\; g^{ij} = \frac{\mdelta^{ij}}{b},
\end{equation}
with all other components vanishing.

Given a solution at temperature $T>0$ to the bulk equations of motion, the fluid-gravity correspondence asserts that given the inhomogeneous boundary conditions above, the bulk fields are: \begin{subequations}\label{eq:EFBG}\begin{align}
\mathrm{d}s^2 &= -2\tilde a(r;H(x))\mathrm{d}v \mathrm{d}r  - a(r;H(x))\mathrm{d}v^2 + b(r;H(x)) \mathrm{d}\mathbf{x}^2 + \mathrm{O}\left(\frac{1}{\xi}\right), \\
\Phi &= \varphi(r;H(x))  + \mathrm{O}\left(\frac{1}{\xi}\right).
\end{align}\end{subequations}
The leading order solution can be found by simply gluing together a locally homogeneous solution.  For our purposes, we can understand this result as follows.  The metric is completely regular near the horizon -- namely, no component of $g_{MN}$ or $g^{MN}$ diverges near the horizon.  Hence, one can check that (for any reasonable norm)\begin{equation}
\left| R_{MN}[g_0] - \frac{R[g_0]}{2}(g_0)_{MN} - T_{MN}^{\mathrm{matter}}[g_0,\Phi_0,\ldots]\right| = \mathrm{O}\left(\frac{1}{\xi}\right),
\end{equation}
where $g_0$, $\Phi_0$ etc. are the metric and matter content given in (\ref{eq:EFBG}).    Furthermore, we can systematically correct $g$ and $\Phi$, order by order in $\xi^{-1}$, in a local fashion.  In contrast to Eddington-Finkelstein coordinates (\ref{eq:EFBG}), the standard Fefferman-Graham coordinates, in which the background metric is diagonal,  have a singular metric at the horizon, and so are not well-suited for the fluid gravity correspondence.

For the purposes of this paper, we will find it sufficient to work with the fluid-gravity expansion at leading order -- the key point is that this expansion is controllable and the perturbative limit $\xi \rightarrow \infty$ is well-behaved.


\section{Charge Diffusion}\label{sec3}
As in \cite{blakeB1} we will compute the charge diffusion constant via the ratio  \begin{equation}
D_{xx} \equiv \frac{\sigma_{xx}}{\chi}.  \label{eq:einstein}
\end{equation}
Here $\sigma$ and $\chi$ are the globally defined electrical conductivity and charge susceptibility, respectively;  these quantities are easier to compute holographically than $D$.   As our striped fluids are anisotropic, both $D$ and $\sigma$ are tensors, but for the remainder of the paper we suppress their explicit indices and focus on diffusion in the striped direction.    (\ref{eq:einstein}) is called an Einstein relation, and its origin is well understood in homogeneous media \cite{kadanoff}.  The common derivation of (\ref{eq:einstein}) in homogeneous media exploits translation invariance and so cannot be applied directly to inhomogeneous media.  Nonetheless, in Appendix \ref{appa}, we show that (\ref{eq:einstein}) remains correct in the inhomogeneous fluids of interest in this paper.    Hence, we  turn to the holographic computation of $\sigma$ and $\chi$.

\subsection{Conductivity}
We begin with the computation of the conductivity $\sigma$.   To accomplish this, we employ the holographic membrane paradigm approach first developed in \cite{iqbal}, and expanded in \cite{donos1409, donos1506, donos1507} to generalize to inhomogeneous geometries.   This technique becomes especially simple in our case, where the translation symmetry is broken only in a single direction \cite{donos1409}.     

In our model, this technique proves especially simple.   The Maxwell equations associated with the bulk gauge field $A$ in (\ref{eq:sec2F2}) are \begin{equation}
\partial_M \left(\sqrt{g} Z(\Phi)F^{MN}\right) = 0.  \label{eq:maxwell}
\end{equation}
To compute $\sigma$, we turn on a perturbation\begin{equation}
A=-Ev\mathrm{d}x + \tilde A.
\end{equation}
The first contribution above imposes a time-independent electric field at the boundary, and the second contribution records the bulk response.   Importantly, since only $F=\mathrm{d}A$ enters (\ref{eq:maxwell}), $\tilde A$ is $v$-independent.   As the background is charge neutral, $A$ will not couple to perturbations of any other bulk fields.   By symmetry, the only possible non-vanishing components of $F$ are $F_{rv}$, $F_{rx}$ and $F_{vx}$, and they are functions of $r$ and $x$ alone.   Hence we may write (\ref{eq:maxwell}) when $N=r$ and $N=x$ respectively as: 
\begin{subequations}
	\begin{align}
	\partial_{x}\left(\sqrt{-g}Z F^{xr}\right)&=0,
	\label{eq:maxwellr}
	\\
	\partial_{r}\left(\sqrt{-g}Z F^{rx}\right)&=0,
	\label{eq:maxwellx}
	\end{align}
\end{subequations}
which immediately implies that \begin{equation}
J = \frac{1}{e^2} \sqrt{-g}Z F^{xr}
\end{equation}
is a constant.   The standard AdS/CFT dictionary may be employed in the UV, and $J$ can be recognized as the expectation value $\langle J_x\rangle$.   From Ohm's Law, \begin{equation}
\sigma = \frac{J}{E}.
\end{equation}
At the horizon, $g^{rr}$ vanishes and so employing our leading order fluid-gravity metric (\ref{eq:EFBG}): \begin{equation}
J = \frac{Z}{e^2} \sqrt{-g} g^{xx}g^{rv} F_{xv} = \frac{Z}{e^2}b^{\frac{d-2}{2}} \left(E-\partial_x \tilde A_v\right).
\end{equation}
We now divide both sides of this equation by the object outside the parentheses, and perform a spatial average, which we denote as $\mathbb{E}[\cdots] = \frac{1}{L_x}\int \mathrm{d}x \cdots$, with $L_x$ denoting the ``length" of the theory in the $x$-direction (possibly infinite).   We find \begin{equation}
\sigma = \dfrac{1}{\mathbb{E}\left[\dfrac{e^2}{Z}b^{-\frac{d-2}{2}}\right]}. \label{eq:sigmafin}
\end{equation}

%
%
%

This equation can be understood through hydrodynamics \cite{lucas}.    The conductivity of a striped fluid with an inhomogeneous local conductivity $\tilde\sigma(x)$ is simply \begin{equation}
\sigma = \dfrac{1}{\displaystyle \mathbb{E}\left[\dfrac{1}{\tilde\sigma(x)}\right]}.  \label{eq:sigmaE}
\end{equation}
The object inside the spatial average in (\ref{eq:sigmafin}) can be interpreted as the inverse conductivity of a fluid, but if it was homogeneous.    In our particular fluid-gravity limit, of course, a hydrodynamic interpretation of $\sigma$ is quite natural, but remarkably a hydrodynamic interpretation persists even beyond the fluid-gravity limit \cite{grozdanov, lucas3, donos1506, donos1507}.

\subsection{Susceptibility}
%
We define the net charge susceptibility as \begin{equation}
\chi \equiv \frac{\partial \mathbb{E}[n]}{\partial \mu },
\end{equation}
where $n = \langle J^t\rangle$ is the charge density of the boundary theory, and $\mu$ is the (spatially homogeneous) chemical potential.   To compute this holographically, we turn on an infinitesimal gauge field $A$ as before, but this time set the boundary conditions in the UV to be \begin{equation}
A(r = 0,x) = \mu \mathrm{d}v.
\end{equation}

In general, there is no elegant membrane paradigm technique to compute $\chi$ in terms of a horizon quantity.   Indeed, as we will see, $\chi$ will depend on details of the full bulk geometry.   Still, let us press ahead, working in a gauge where $A_r=0$.   The $v$-component of (\ref{eq:maxwell}) reads \begin{equation}
\partial_r \left( \sqrt{-g} Z F^{rv}\right)  + \partial_x \left( \sqrt{-g} Z F^{xv}\right) = 0. \label{eq:32max}
\end{equation}
As we will justify more carefully soon,  the second term can be neglected in the fluid-gravity limit.  Hence, in our gauge, we obtain an ``effective" radially conserved quantity at each $x$ \begin{equation}
n = -\frac{1}{e^2} \sqrt{-g} Z F^{vr} =- \frac{1}{e^2} b^{\frac{d}{2}} \tilde a^{-1} Z \partial_r A_v,  \label{eq:naV}
\end{equation}
which the AdS/CFT dictionary identifies as the boundary theory expectation value of $\langle J^t\rangle$.   This justifies why we have labeled this radially conserved quantity $n$, as it is the local charge density.   Now, we employ the following trick \cite{blakeB1}: if the horizon is located at $r=r_+$, since $A_v(r=r_+)=0$ in order for $A$ to be well-defined, \begin{equation}
n(x) \int\limits_0^{r_+} \mathrm{d}r \frac{e^2 \tilde a}{Zb^{\frac{d}{2}}}  = \int\limits_0^{r_+}  \mathrm{d}r \left(-\partial_r A_v\right) = \mu,  \label{eq:nxmu}
\end{equation}
where we have employed the boundary conditions in the last step.  We note that $b\sim \tilde a \sim r^{-2}$ as $r\rightarrow 0$, and that both $b$ and $\tilde a$ are regular at the horizon; thus $A_v$ will be finite everywhere.  Hence, upon applying the uniform chemical potential $\mu$, we see that \begin{equation}
\chi  = \frac{\partial \mathbb{E}[n]}{\partial \mu} = \mathbb{E}\left[ \left(\int\limits_0^{r_+} \mathrm{d}r \frac{e^2 \tilde a}{Zb^{\frac{d}{2}}}\right)^{-1}\right].  \label{eq:chifin}
\end{equation}

To confirm that this is the correct answer at leading order in $\xi^{-1}$, we note that (\ref{eq:naV}), together with the value for $n(x)$ in (\ref{eq:nxmu}), gives us an approximate solution for $A_v$ at all $r$ and $x$, which is smooth everywhere and will not exhibit any singular behavior at finite $T$.        By simply plugging this ``guess" into (\ref{eq:32max}), we see that it is correct to leading order:   regularity of $A_v$, and the geometry, ensures that all corrections in $\xi^{-1}$ are subleading.

In the limit of hydrodynamic disorder, because $n(\mu,H,\ldots)$ is a local quantity we expect that \begin{equation}
\chi = \mathbb{E}\left[\tilde\chi(x)\right], \label{eq:chiE}
\end{equation}
with $\tilde \chi(x) = \partial n(x)/\partial \mu$ a local susceptibility.   Our holographic computation in the fluid-gravity geometries confirms this explicitly.   
%
%

To summarize, the diffusion coefficient can now be written in terms of spatial averages as 
\begin{equation}
D=\frac{1}{\mathbb{E}\Big[b^{\frac{2-d}{2}}Z(r;H(x))^{-1}\Big]\mathbb{E}\Big[\Big(\int_{0}^{r_+}\mathrm{d}r b^{-\frac{d}{2}}\tilde{a}Z(r;H(x))^{-1}\Big)^{-1}\Big]}.   \label{eq:Dend3}
\end{equation}
This result can essentially be understood through classical hydrodynamics.   The reason that it is nonetheless useful to do the computation holographically is that we now have an explicit formula for $D$ in terms of bulk quantities.  It is not obvious that the butterfly velocity $v_{\textsc{b}}$ -- which by definition probes \emph{quantum} chaos -- admits any classical interpretation.   We will now be able to compare $v_{\textsc{b}}$ to $D$.

\section{Butterfly Velocity}\label{sec4}
Recall that the butterfly effect is a phenomena in chaotic systems in which an initially small perturbation can grow exponentially large at later times.   In the introduction, we captured this in terms of a peculiar 4-point correlation function (\ref{eq:sec1vb}).   Several recent works have pointed out that the holographic interpretation of the butterfly effect is a geometric shock wave, propagating along the horizon \cite{bhbutterfly,localized}.    We will not fully justify why here, only explain the basic idea.   Black holes such as the ones we have constructed can be maximally extended to ``double-sided" geometries, which contain two copies of the field theory, where time runs in opposite directions, along with a black hole and a white hole \cite{wald}.     This maximally extended geometry can be interpreted as two entangled copies of a field theory in a particular excited and entangled state.    We now imagine slightly perturbing the left field theory (where we take time to run backwards) at a ``late" time $t=t_0$, by adding a small amount of massless particles.   This should set off the butterfly effect, and the geometry at time $t=0$ should be very different than what it was without the perturbation.   

To leading order, we might expect that this perturbation does nothing to the classical geometry in the limit $G_{\mathrm{N}}\rightarrow 0$.   However, if $t_0$ is large, then these massless particles will follow null trajectories very close to the (past) horizon, where they become exponentially blue-shifted,  analogous to particles falling into the classic Schwarzchild black hole.   We then must solve Einstein's equations in the presence of a thin shell of energy associated with this in-falling blue-shifted matter \cite{masslessshock,curvedshock}.   The resulting geometry will contain a shock at the horizon (a sudden shift in the geometry), which allows us to capture both $\lambda$ and $v_{\textsc{b}}$, as defined in (\ref{eq:sec1vb}).

\subsection{Kruskal Coordinates}
\begin{figure}
\centering
\includegraphics[width=5in]{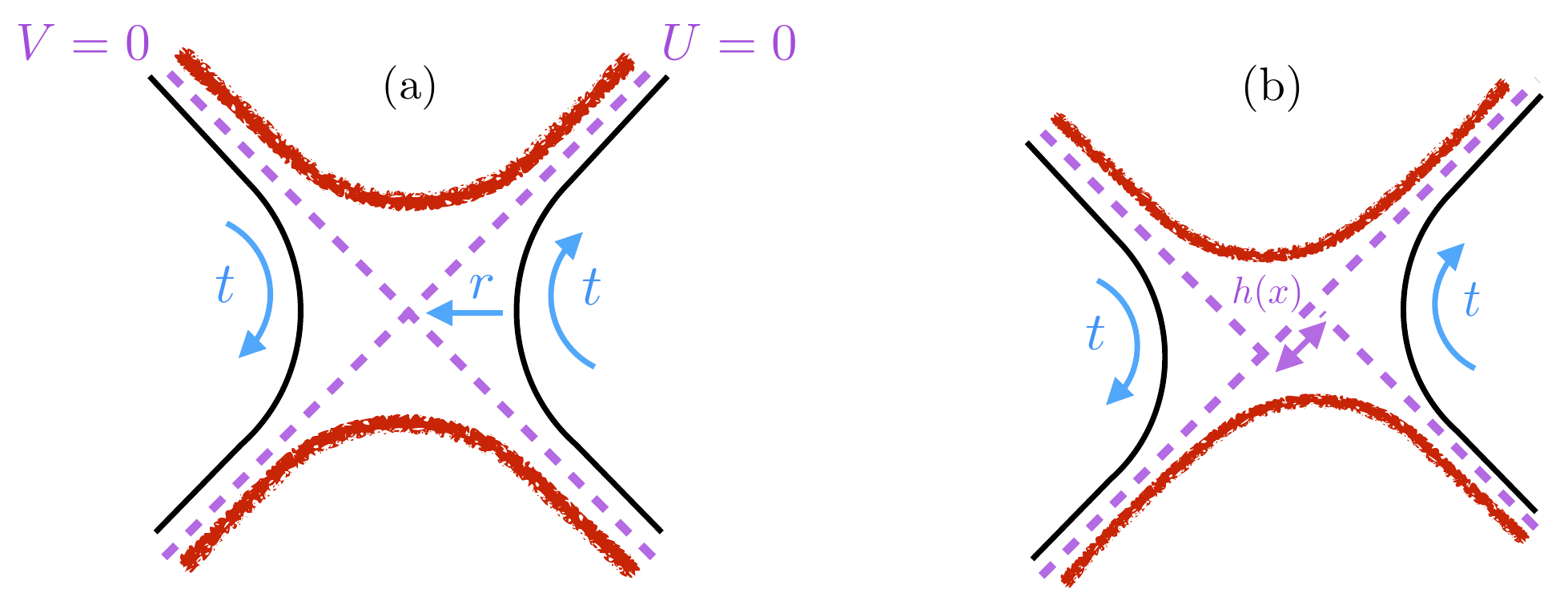}
\caption{(a) The global Kruskal geometry.   The smooth black lines on the left/right side of the diagram denote the ``AdS" boundaries of the left/right field theories.   The rough red lines at the top and bottom denote the singularities inside of a black hole and a white hole respectively.   The dashed purple lines denote event horizons at $U=0$ and $V=0$.   Blue arrows denote the direction as well as the standard ``bulk radial" coordinate $r$.   Eddington-Finkelstein coordinates only cover the region $V\ge 0$ in this diagram.   The properties of this black hole vary smoothly in the $x$ direction, which runs out of the page.  (b) The effect of the ``geometric shockwave" is a shift $h(x)$ between the horizons as one crosses $U=0$.}
\label{fig1}
\end{figure}

We have now outlined the computation of $v_{\textsc{b}}$ in words.   The first thing to do is now to adopt a coordinate system which completely covers the maximally extended black holes of interest, called Kruskal coordinates.  For simplicity, we begin with homogeneous black holes (corresponding to the right side field theory in the maximally extended case), where \begin{equation}
		\mathrm{d}s^2 = 2A(UV; H)\mathrm{d}U \mathrm{d}V +B(UV;H) \mathrm{d}\mathbf{x}^2 .
\end{equation}
In terms of the null in-going coordinate $v$, and a null out-going coordinate \begin{equation}
\mathrm{d}u \equiv -\frac{2\tilde a}{ a} \mathrm{d}r - \mathrm{d}v,
\end{equation}the Kruskal coordinates $U$ and $V$ are defined as \begin{subequations}\begin{align}
U &= L\mathrm{e}^{2\mpi T u} , \\
V &= L \mathrm{e}^{2\mpi Tv}.
\end{align}\end{subequations}
Hence, \begin{subequations}\label{eq:kruskalAB}\begin{align}
A &= \frac{2a}{(4\mpi T)^2 UV}, \\
B &= b.
\end{align}\end{subequations}
The event horizons are located at $U=0$ and/or $V=0$.  Note that $A$ and $B$ are functions only of the product $UV$, and that despite appearances, $A$ is in fact regular at $UV=0$.   Figure \ref{fig1}(a) shows the global spacetime in Kruskal coordinates.

Now, we return to the case where $H(x)$ is not a constant.   In fact, to leading order in $\xi^{-1}$, the inhomogeneous black hole in Kruskal coordinates is given by: \begin{subequations}\label{eq:kruskalBG}\begin{align}
\mathrm{d}s^2 &= 2A(UV; H(x))\mathrm{d}U \mathrm{d}V +B(UV;H(x)) \mathrm{d}\mathbf{x}^2  + \mathrm{O}\left(\frac{1}{\xi^2}\right), \\
\Phi &=  \phi(UV;H(x)).
\end{align}\end{subequations}
We justify this claim more carefully in Appendix \ref{app:flugrav}.  Kruskal coordinates are not commonly employed for a ``fluid-gravity" correspondence.   In Eddington-Finkelstein coordinates, given \emph{time-dependent} fluid flows in the boundary theory, the bulk geometry may be constructed ``tube-wise":  namely, given data at a boundary spacetime point $(t,\mathbf{x})$,  one can construct the local geometry for all $r$.    In Kruskal coordinates, $U$ and $V$ both relate to time $t$, and so one would need dynamics for all $t$ to construct the bulk geometry on for any given $\mathbf{x}$.   However, since we are looking at static geometries, we trivially know the boundary ``fluid dynamics" for all times $t$.    Furthermore, Kruskal coordinates share the same key regularity property of Eddington-Finkelstein coordinates:  both $g_{MN}$ and $g^{MN}$ are completely regular at $U=0$ or $V=0$.   Thus, it is guaranteed that (\ref{eq:kruskalBG}) indeed solves the equations of motion at leading order in $1/\xi$, and that the effects of perturbations are not singular.  Hence, we can use a \emph{local} fluid-gravity expansion to construct an inhomogeneous static black hole, order-by-order in $1/\xi$, directly in Kruskal coordinates.   We explicitly discuss the construction in Appendix \ref{app:flugrav}.   Our main result can be obtained  from the leading order solution (\ref{eq:kruskalBG}).

\subsection{Shockwave Geometry}
As we mentioned previously, we consider a small amount of matter thrown into the left-half of our Kruskal black hole at an ``early" time $t=t_0$.    For simplicity, we consider a thin shell of matter, located at $x=0$, and uniform in the $d-1$ spatial directions perpendicular to $x$.    As this matter falls towards the horizon, it creates a very sharp distribution of energy which we approximate as \begin{equation}
T_{UU}^{\mathrm{shock}} = \mathcal{E} \mathrm{e}^{2\mpi Tt} \mdelta(x)\mdelta(U) ;
\end{equation}
all other components of the stress tensor vanish.     The $\mdelta(x)$ is not a true $\mdelta$ function, but can be approximated as such for studying long distance physics \cite{localized} -- at the end of this section, we will provide a few more comments on this assumption.    Here $\mathcal{E}$ is the (proper) energy density of the infalling shell of matter, and $\mathrm{e}^{2\mpi Tt}$ is a blue-shift factor as it falls towards the horizon.
The solution to Einstein's equations in the background of such a source has been known for some time in a homogeneous geometry \cite{curvedshock}.   The presence of this pulse of energy located at the $U=0$ horizon creates a mismatch in the location of the horizon as one passes from the left to the right side (see Figure \ref{fig1}(b)):  
\begin{equation}
\label{eq:shock}	V \rightarrow V+\mathrm{\Theta}(U) h(x), \enspace \enspace\enspace  U\rightarrow U, \enspace \enspace \enspace\mathbf{x} \rightarrow \mathbf{x}.
\end{equation}
Using (\ref{eq:shock}), we see this results in the following backreacted metric
\begin{equation} \label{eq:shockwavemetric0}
\mathrm{d}s^2 = 2A(UV;H(x))\mathrm{d}U \mathrm{d}V +B(UV;H(x)) \mathrm{d}\mathbf{x}^2 -2A(UV;H(x))\delta(U)\mathrm{d}U^{2}+ \mathrm{O}\left(\frac{1}{\xi}\right).
\end{equation}
The stress tensor is modified to
\begin{equation}
T_{MN}=T^{(0)}_{MN}+T^{(1)}_{MN}
\end{equation}
where $T^{(0)}_{MN}$ is the stress tensor without the shock wave, and the only non-zero component of $T^{(1)}_{MN}$ is (at leading order in $\xi^{-1}$)
\begin{equation}
	T^{(1)}_{UU}=\mathcal{E}\mathrm{e}^{2\mpi Tt_0}\mdelta(U)\mdelta(x)-2h(x,t)\mdelta(U)T^{0}_{UV}.
	\end{equation}
We have used the fact that on the background, $T^{(0)}_{VV}=0$ \cite{curvedshock} in deriving this result.  We discuss higher order corrections to this shock wave geometry in Appendix \ref{app:vbcorrections}.

We now follow and compute $v_{\textsc{b}}$ by solving Einstein's equations for $h(x)$.   This computation -- in the homogeneous case $H(x)=\text{constant}$ -- is well-understood \cite{curvedshock, blakeB1, lrbutterfly}, and so we focus on the new effects coming from inhomogeneity.   It is instructive to see how the shock wave affects the Ricci tensor $R_{MN}$.   The only shock-dependent contributions to $R_{MN}$ coming from the metric (\ref{eq:shockwavemetric0}) are in the $UU$ component, (keep in mind that further corrections are possible at $\mathrm{O}(\xi^{-2})$ due to fluid-gravity corrections to (\ref{eq:shockwavemetric0}), as discussed in Appendix \ref{app:vbcorrections}): \begin{equation}
R^{(1)}_{UU} = \mdelta(U) \left[\frac{A}{B} \partial_x^2 h + \frac{1}{B} \partial_x (\partial_x A h) + \frac{d-2}{2}\frac{\partial_x B}{B^2}\partial_x (Ah) +2h\frac{\partial_{UV}A}{A} + \frac{d}{2}h\frac{\partial_{UV}B}{B} \right].  \label{eq:47}
\end{equation}
To simplify these results, we have used the following equalities, which hold in the sense of distributions:\footnote{A transparent way to see the last equality is to approximate $\mdelta(u) = \epsilon^{-1}f(u/\epsilon)$ with $\epsilon \rightarrow 0$ and $f$ a suitably chosen function:  e.g. $f(x)=\frac{1}{2}\mathrm{e}^{-|x|}$.   One finds $\int \mathrm{d}u (u \mdelta(u))^2 \sim \epsilon$ which vanishes in the $\epsilon \rightarrow 0$ limit.}
 \begin{equation}
U\mdelta(U) = U^2\mdelta(U) = U^2\mdelta(U)^2 = 0, \;\;\;\; \mdelta(U) = -U\mdelta^\prime(U).  \label{eq:UDelta}
\end{equation}
From (\ref{eq:47}) we see that $x$-derivatives do not enter the expression for $R^{(1)}_{UU}$ in a singular way near $U=0$; similar calculations confirm this result for the remainder of Einstein's equations.  Assuming that the matter content is also well-behaved (we give an explicit example where this is so in Appendix \ref{app:vbcorrections}), we may take the $\xi \rightarrow \infty$ limit smoothly.   Using that, at order $\xi^0$, the background Einstein equations give \cite{curvedshock} \begin{equation}
8\mpi G_{\mathrm{N}} T_{UV}^{(0)} = \frac{\partial_{UV}A}{A} + \frac{d}{2}\frac{\partial_{UV}B}{B},
\end{equation}
one can show that the coefficient of $\mdelta(U)$ in the $UU$ component of Einstein's equations (with lowered indices) only vanishes when $h(x)$ obeys the differential equation \cite{curvedshock} \begin{equation}
\left(\partial_x^2 - m(x)^2\right)h(x) = 8\mpi G_{\mathrm{N}} \frac{B(0,0)}{A(0,0)} \mdelta(x) \mathcal{E}\mathrm{e}^{2\mpi Tt},
\end{equation}
with the ``mass" $m$ is defined as
\begin{equation}
m(x)^2\equiv \frac{d}{2A(0;H(x))}\partial_{UV}B(UV;H(x))\Big\vert_{U=0}. \label{eq:butterfly}
\end{equation}

We now solve for the profile $h(x)$.   This can be done using a standard WKB-like argument.   Let us define \begin{equation}
h(x) \equiv \exp\left[-\left|\int\limits_0^x \mathrm{d}x^\prime \; m(x^\prime)\right|\right] h_0(x),\label{eq:h0x}
\end{equation}
where $h_0(x)$ obeys the differential equation \begin{equation}
\partial_x^2 h_0 - 2m(x) \partial_x h_0  -  h_0 \partial_x m(x)  = 8\mpi G_{\mathrm{N}} \frac{B(0,0)}{A(0,0)} \mdelta(x) \mathcal{E}\mathrm{e}^{2\mpi Tt}.  \label{eq:h0x2}
\end{equation}  
If we demand, as is physically sensible, that $h(x)$ vanish at $x=\pm \infty$, then (\ref{eq:h0x2}) implies that $\partial_x h_0 = \mathrm{O}(\xi^{-1})$.   Thus, we conclude that at leading order in $\xi$, we may approximate (at large $x$):  \begin{equation}
h \approx 8\mpi G_{\mathrm{N}} \frac{B(0,0)}{A(0,0)} \mathcal{E}\mathrm{e}^{\lambda(t-|x|/v_{\textsc{b}})},
\end{equation}
with $\lambda $ given by (\ref{eq:lambda}) and \begin{equation}
\frac{1}{v_{\textsc{b}}} \equiv \mathbb{E}\left[\frac{1}{\tilde v_{\textsc{b}}}\right]  \label{eq:1vb}
\end{equation}
with \begin{equation}
\frac{1}{\tilde v_{\textsc{b}}} \equiv \frac{1}{2\mpi T} \sqrt{\frac{d}{2A(0;H(x))}\partial_{UV}B(0;H(x))}
\end{equation}

Interestingly, the butterfly velocity in this inhomogeneous background can be interpreted in a very simple classical picture.   Imagine a ``butterfly"  moving through a medium with local velocity $\tilde v_{\textsc{b}}(x)$.    What would the effective velocity of the butterfly be?   It is easiest to compute this effective velocity $v_{\textsc{b}}$ by measuring the time it takes to travel a distance $\ell$:  \begin{equation}
t(\ell) = \int \mathrm{d}t = \int \frac{\mathrm{d}x}{\tilde v_{\textsc{b}}(x)} = \ell \; \mathbb{E}\left[\frac{1}{\tilde v_{\textsc{b}}}\right] \equiv \frac{\ell}{v_{\textsc{b}}},
\end{equation}
in agreement with our (leading order) holographic calculation (\ref{eq:1vb}).   Although this simple classial picture is intuitive, it is worth keeping in mind that the butterfly velocity is a manifestation of \emph{quantum} chaos.

The derivation above assumed that the influence of the ``pulse" at the horizon was localized.   However, our derivation of the butterfly velocity is not particularly sensitive to this assumption.  As we have seen, $h(x)$ obeys a linear differential equation.  Even if $h(x)$ is sourced by a source of finite width $a$,   for distances $x\gg a$,  $h(x)\sim \mathrm{e}^{-\lambda x/v_{\textsc{b}}}$ by linearity, and so our derivation of the butterfly velocity is unchanged.

\section{Scaling Geometries}\label{sec5}
We now compare $D$ and $v_{\textsc{b}}$.   It is instructive to begin with a simple Einstein-Maxwell-dilaton (EMD) holographic models, with bulk action \begin{equation}
S = \int \mathrm{d}^{d+2}x \sqrt{-g} \left[\frac{1}{16\mpi G_{\mathrm{N}}}\left(R - \frac{1}{2}(\partial \Phi)^2 - \mathcal{V}(\Phi)\right) - \frac{Z(\Phi)}{4}F^2\right].  \label{eq:sec5action}
\end{equation}
We have carefully checked the validity of the fluid-gravity expansion in Kruskal coordinates in Appendix \ref{app:flugrav}, as well as subleading corrections to $v_{\textsc{b}}$ in Appendix \ref{app:vbcorrections}, for this model.  

Let us now consider the low-temperature scaling limit of our gravity theory  \cite{charmousis, rmp}.   Let us temporarily assume that $H(x)=\text{constant}$.   For ``generic" Liouville-like potentials: \begin{equation}
\mathcal{V}(\Phi \rightarrow \infty) = \mathcal{V}_0 \mathrm{e}^{\beta\Phi},
\end{equation}   
one finds, at $T=0$, the emergent IR geometry (in Eddington-Finkelstein coordinates, in a certain gauge) is a hyperscaling-violationg geometry:  \begin{equation}
\mathrm{d}s^2 = \frac{L^2}{r^2}\left(H^{\frac{1}{d+1-\Delta}}r\right)^{\frac{2\theta}{d}} c_0 \left(2\mathrm{d}v\mathrm{d}r - \mathrm{d}v^2 +   \mathrm{d}\mathbf{x}^2\right) ,  \label{eq:sec5metric0}
\end{equation}
so long as $\theta<0$ (or $\theta>d$, though this case may be unphysical \cite{swingle}).   This requirement follows from the formula \cite{charmousis, rmp} \begin{equation}
\beta^2 =  \frac{2\theta}{d(d-\theta)},
\end{equation}
and the fact that $\beta$ must be real.    We emphasize that (\ref{eq:sec5metric0}) is not the full geometry, and that this approximation will fail for $r\lesssim r_{\mathrm{uv}} \sim H^{\frac{1}{d+1-\Delta}}$ -- for smaller values of $r$, there is a UV completion to an asymptotically AdS geometry, the details of which are unimportant to us.    At low temperatures \begin{equation}
 \frac{T}{H^{\frac{1}{d+1-\Delta}}} \ll 1,   \label{eq:THlim}
 \end{equation} the IR geometry is only modified from (\ref{eq:sec5metric0}) in a simple manner.   The IR emergent hyperscaling geometry remains, but now with a planar black hole horizon.   The metric is approximately given by \begin{equation}
\mathrm{d}s^2 = \frac{L^2}{r^2}\left(H^{\frac{1}{d+1-\Delta}}r\right)^{\frac{2\theta}{d}} c_0 \left(2\mathrm{d}v\mathrm{d}r - f_T(r)\mathrm{d}v^2 +   \mathrm{d}\mathbf{x}^2\right) , \label{eq:sec5metric}
\end{equation}
with emblackening factor \begin{equation}
f_T(r) = 1- \left(\frac{r}{r_+}\right)^{d+1-\theta},
\end{equation}
where \begin{equation}
r_+ =\frac{d+1-\theta}{4\mpi T}.
\end{equation}
 In this geometry, $\Phi \sim \log r$ in the IR scaling region.   Hence, we expect, for canonical choices $Z(\Phi) \sim \exp[\beta^\prime \Phi]$, that \begin{equation}
Z = Z_0 \left(H^{\frac{1}{d+1-\Delta}}r\right)^\gamma,
\end{equation}
for some scaling exponent $\gamma$.    Again, this IR geometry is only valid for $r\lesssim r_{\mathrm{uv}}$.

Of course, we wish to study geometries where $H(x)$ is not constant.   However, as we have discussed in Section \ref{sec2},  so long as (\ref{eq:fglim}) is satisfied, these geometries can be constructed by simply ``gluing" together the homogeneous geometries point-by-point in $x$, using the local value for $H(x)$ in (\ref{eq:sec5metric}).   It is critical that (\ref{eq:fglim}) is obeyed -- if we take the limit $T\rightarrow 0$ with $\xi$ fixed, then we expect such a perturbation to decay in the IR geometry (see e.g. \cite{chesler}).    Because we further assume (\ref{eq:THlim}), the homogeneous geometries which we must glue together take a particularly simple form (\ref{eq:sec5metric}) in the IR, at leading order in $1/\xi T$ and $T/H^{\frac{1}{d+1-\Delta}}$, point-by-point in $x$.    We also remind the reader that the validity of the fluid-gravity expansion is not sensitive to the assumption (\ref{eq:THlim}) -- indeed, (\ref{eq:EFBG}) holds more generally so long as (\ref{eq:fglim}) holds.   Nevertheless, we will be interested in geometries where the near-horizon limit of metrics such as (\ref{eq:EFBG}) can be approximated by (\ref{eq:sec5metric}), with an $x$-dependent $H$.

Using the results of the previous sections, it is simple to compute $D$ and $v_{\textsc{b}}$.   We begin by computing $\sigma$, $\chi$ and $D$.   Using (\ref{eq:sigmafin}), along with the near-horizon geometry (\ref{eq:sec5metric}) we obtain a local effective conductivity\begin{equation}
\tilde \sigma = \frac{1}{e^2}b(r_+)^{\frac{d}{2}-1} Z(r_+) = \frac{Z_0}{e^2} c_0^{\frac{d}{2}-1} H^{\frac{1}{d+1-\Delta}(\frac{2\theta}{d}(\frac{d}{2}-1) + \gamma)}   r_+^{\gamma + (\frac{2\theta}{d}-2)(\frac{d}{2}-1)}.   \label{eq:sec5sigma}
\end{equation} Next, we use (\ref{eq:chifin}) to compute $\tilde \chi$.   At $T\sim H^{\frac{1}{d+1-\Delta}}$, $\tilde \chi$ will be a complicated function of $T$:  \begin{equation}
\tilde \chi \approx \left(\int\limits_{r_{\mathrm{uv}}}^{r_+} \frac{e^2 \mathrm{d}r }{Z_0 c_0^{\frac{d}{2}-1} H^{\frac{1}{d+1-\Delta}(\frac{2\theta}{d}(\frac{d}{2}-1) + \gamma)}  r^{\gamma + (\frac{2\theta}{d}-2)(\frac{d}{2}-1)}}  + \int\limits_0^{r_{\mathrm{uv}}} \mathrm{d}r \frac{e^2 \tilde a}{b^{d/2} Z}  \right)^{-1} \label{eq:chi05}
\end{equation}
Recall $r_{\mathrm{uv}}$ is the scale at which (\ref{eq:sec5metric}) fails to be a good approximation.   Since this scale is set by $H$ and not $T$, in the limit (\ref{eq:THlim}), we note that only the first term of (\ref{eq:chi05}) is $T$-dependent.   We focus on the limit where this is the dominant term in $\tilde \chi$ in the limit $r_+\gg r_{\mathrm{uv}}$,\footnote{If this term is subleading, then $\tilde \chi$ will be sensitive to the UV geometry, whereas $\tilde \sigma$ and $\tilde v_{\textsc{b}}$ both depend only on the near-horizon geometry.   Hence there can be no universal relation between $D$ and $v_{\textsc{b}}$, a point noted in \cite{blakeB1}.} which occurs when \begin{equation}
\left(1-\frac{\theta}{d}\right)\left(d-2\right)-\gamma> -1
\end{equation}
 In \cite{blakeB1}, it was noted that this will happen if the IR scaling dimension for the charge susceptibility is positive.\footnote{This assumes that we interpret generic exponents $\gamma$ as anomalous dimensions for the charge density, as in \cite{blaise1308, hartnollscale}.}   In this case,   (\ref{eq:chi05}) can be approximated as   \begin{align}
\tilde \chi &= \left(1+2\left(1-\frac{\theta}{d}\right)\left(\frac{d}{2}-1\right) - \gamma\right) Z_0 c_0^{\frac{d}{2}-1} H^{\frac{1}{d+1-\Delta}(\frac{2\theta}{d}(\frac{d}{2}-1) + \gamma)} r_+^{\gamma -1 + (\frac{2\theta}{d}-2)(\frac{d}{2}-1)}  \notag \\
&= \left(1+2\left(1-\frac{\theta}{d}\right)\left(\frac{d}{2}-1\right) - \gamma\right) \tilde \sigma \frac{4\mpi T}{d+1-\theta}.  \label{eq:sec5chi}
\end{align}
Combining (\ref{eq:Dend3}), (\ref{eq:sec5sigma}) and (\ref{eq:sec5chi}) we obtain $D$: \begin{equation}
D = \frac{d+1-\theta}{1+2(1-\frac{\theta}{d})(\frac{d}{2}-1)-\gamma} \frac{1}{4\mpi T \mathbb{E}[H^\eta]\mathbb{E}[H^{-\eta}]}
\end{equation} 
where the power $\eta$ is generically not zero:\begin{equation}
\eta =  \frac{1}{d+1-\Delta}\left(\frac{\theta}{d}(d-2) + \gamma\right).
\end{equation}
 If the integral for $\chi$ is dominated away from the horizon,  then the diffusion constant becomes parametrically large \cite{blakeB1}.

Let us now turn to the computation of $v_{\textsc{b}}$.   Using (\ref{eq:kruskalAB}), we find that \begin{equation}
\frac{1}{v_{\textsc{b}}^2} = \frac{(4\mpi T)^2 dUV}{(2\mpi T)^2 4a} \partial_{UV}b \approx \frac{(r_+-r)d}{a}  (-\partial_r b).
\end{equation}
In the latter step, we have taken the near-horizon limit.  For the geometries (\ref{eq:sec5metric}), this leads to the simple constant result:
\begin{equation}
\tilde v_{\textsc{b}}^2 = \frac{d+1-\theta}{2(d-\theta)},
\end{equation}
in agreement with \cite{blakeB1, lrbutterfly}, though now in more generic striped backgrounds.   Interestingly, in this special case where $v_{\textsc{b}} \sim T^0$, this velocity has also been found to characterize entanglement entropy growth during thermalization \cite{alishahiha}.   

If we assume that $H$ is a constant, we find that the constant $\mathcal{C}$ defined in (\ref{eq:diffLB3}) is \begin{equation}
\mathcal{C} = \frac{d-\theta}{2(1-\frac{\theta}{d})(\frac{d}{2}-1)-\gamma}.
\end{equation}
However, if $H(x)$ is not constant, then $D$ is not related to $v_{\textsc{b}}$ in a simple manner, regardless of the value of $\gamma$.   Using the Cauchy-Schwarz inequality (see e.g. \cite{ikeda}), we find that \begin{equation}
\mathbb{E}\left[H^\eta\right]\mathbb{E}\left[H^{-\eta}\right] \ge 1.
\end{equation}
Hence, we conclude that the inequality of (\ref{eq:diffLB3}) holds:  namely, $2\mpi T D \le \mathcal{C} v_{\textsc{b}}^2$.   In principle, there is no bound on the ratio $D v_{\textsc{b}}^{-2}$ -- the diffusion constant can be parametrically small compared to the butterfly velocity.     As we emphasized in the introduction, this inequality is the ``wrong sign" -- an incoherent metal should have diffusion constants bounded from below, and so the butterfly velocity evidently cannot always serve as the velocity scale in (\ref{eq:diffLB2}) in any sharp sense.   However, (\ref{eq:diffLB2}) continues to hold in a qualitative sense, as in these simple models $v_{\textsc{b}}\sim T^0$ and $D\sim 1/T$, in agreement with the scaling noted in \cite{blakeB1}.

We also expect our conclusions to hold in more complicated bulk models where we have not explicitly checked that the fluid-gravity expansion is well-behaved in Kruskal coordinates.   In more general scaling regimes where the bounds of \cite{blakeB1} hold in homogeneous systems, we find \emph{locally} \begin{equation}
\frac{\tilde \sigma}{\tilde \chi} = \frac{\mathcal{C}}{2\mpi T} \tilde v_{\textsc{b}}^2.
\end{equation}
Then \begin{equation}
\frac{D}{v_{\textsc{b}}^2} \frac{2\mpi T}{\mathcal{C}} = \frac{\mathbb{E}[\tilde v_{\textsc{b}}^{-1}]^2}{\mathbb{E}[\tilde \chi]\mathbb{E}[\tilde \sigma^{-1}]} \frac{2\mpi T}{\mathcal{C}}  = \frac{\mathbb{E}[ \tilde\sigma^{-1/2}\tilde\chi^{1/2}]^2}{\mathbb{E}[\tilde \chi]\mathbb{E}[\tilde \sigma^{-1}]} \le 1,
\end{equation}
again by the Cauchy-Schwarz inequality, as $\tilde \sigma$ and $\tilde \chi$ are positive everywhere.   Our result that the charge diffusion constant is upper bounded by the butterfly velocity in these striped models is not peculiar to the Einstein-dilaton model we studied above.   The inequality (\ref{eq:diffLB3}) may change beyond the hydrodynamic limit, or when the disorder breaks translation symmetry in multiple directions.

\section{Conclusions}
In this paper, we have shown that in striped charge neutral holographic quantum matter, the butterfly velocity cannot generally be used to quantitatively provide a lower bound for the charge diffusion constant.   This implies that, to the extent that (\ref{eq:diffLB2}) should hold exactly in any incoherent metal,  the velocity scale $v$ in (\ref{eq:diffLB2}) either cannot be the butterfly velocity.   It would be interesting if there is a slower velocity scale for which (\ref{eq:diffLB2}) remains true, even in the striped geometries we have constructed in this paper.   We also note that we did not present an explicit example where $DT/v_{\textsc{b}}^2$ had non-trivial $T$-dependence as $T\rightarrow 0$, though we cannot rule this possibility out.

In general, bounds on transport coefficients, which are robust to the specific nature of disorder, are known to exist only in a handful of holographic systems \cite{grozdanov, grozdanov2, ikeda}.   It is always the case that complicating the bulk models sufficiently can lead to the violation of any naive bound.   Similarly, a $v_{\textsc{b}}$-based bound on charge diffusion might hold in special models, even if it does not hold more generally.  It would be interesting to more precisely determine what such cases are.    Finally, we comment that in this charge-neutral hydrodynamic limit,  although charge diffusion is very weak,  the energy diffusion constant will be very large \cite{donos1507}.    It may be the case that, similarly to \cite{grozdanov2}, there are holographic models where a $v_{\textsc{b}}$-based bound on energy diffusion is very robust.  As both $D$ and $v_{\textsc{b}}$ do not seem bounded in the incoherent limit in ``mean-field" disordered holographic metals \cite{blakeB2}, it would be interesting to explore this further in future work.

Even beyond the hydrodynamic limit, the direct current transport coefficients of most holographic models can be computed in terms of an emergent hydrodynamic on the black hole horizon (at finite temperature)  \cite{donos1409, donos1506, donos1507, lucas}.   It would be interesting if, for arbitrary black holes, there was some simple partial differential equation governing the shift function $h$, defined in Section \ref{sec4}.   This may lead to fundamental bounds on the butterfly velocity, analogously to how conductivity bounds may be found \cite{grozdanov, grozdanov2, ikeda}.
\addcontentsline{toc}{section}{Acknowledgements}
\section*{Acknowledgements}
We thank Mike Blake, Sean Hartnoll and Subir Sachdev for helpful discussions.   A.L. was supported by the NSF under Grant DMR-1360789 and MURI grant W911NF-14-1-0003 from ARO.   J.S. was supported by the National Science Foundation Graduate Research Fellowship under Grant No. DGE1144152.
\begin{appendix}

\section{Diffusion in an Inhomogeneous Fluid}\label{appa}
There is a rather controversial history of Einstein relations in inhomogeneous media (in many cases this involves situations with inhomogeneous temperature \cite{vankampen}).   In this appendix, we directly confirm that the Einstein relation employed in the main text is correct in the hydrodynamic limit, suitable for our holographic computation.

As in the main text, we consider a background charge neutral fluid, and study the linearized propagation of charge fluctuations around this background.   The equations of hydrodynamics simplify to ``resistor network" equations (in the continuum) \cite{lucas}:\begin{equation}
\partial_t n  = \partial_x \left(\tilde\sigma(x) \partial_x \left(\frac{n}{\tilde \chi(x)}\right)\right),  \label{eq:Adiff}
\end{equation}
where $\tilde \sigma$ is the local conductivity of the fluid (a dissipative coefficient within hydrodynamics) and $\tilde \chi$ is the local charge susceptibility,  which are related to the global $\sigma$ and $\chi$ through (\ref{eq:sigmaE}) and (\ref{eq:chiE}).    Our goal is now to show that on long time and length scales, in some suitable sense, \begin{equation}
n(x,t) \sim \mathrm{e}^{\mathrm{i}qx - Dq^2t}
\end{equation}
is a ``solution" to (\ref{eq:Adiff}), with $D$ given by (\ref{eq:einstein}).     More precisely, we show that there is an asymptotic solution to (\ref{eq:Adiff}) of the form \begin{equation}
n(x,t) = \mathrm{e}^{\mathrm{i}qx - Dq^2t} \left[b_0(x) + \mathrm{i}q b_1(x) - q^2 b_2(x) + \cdots\right].  \label{eq:nxt}
\end{equation}

By comparing (\ref{eq:nxt}) into (\ref{eq:Adiff}) at $q=0$, we start with a simple solution\begin{equation}
b_0 = \tilde\chi.
\end{equation}
At $\mathrm{O}(q)$, we find that (\ref{eq:Adiff}) reads \begin{equation}
0 = \partial_x \left(\tilde \sigma \partial_x \left(\frac{b_1}{\chi}\right) + \tilde \sigma \right).
\end{equation}
This equation is readily integrated: \begin{equation}
\partial_x \left( \frac{b_1}{\tilde \chi}\right) = -1 + \frac{C}{\tilde \sigma},  \label{eq:1C0}
\end{equation}
with $C$ an integration constant.   We fix $C$ by demanding that $b_1$ not diverge with $x$: \begin{equation}
\frac{1}{C}= \mathbb{E}\left[\frac{1}{\tilde\sigma}\right].  \label{eq:1C}
\end{equation}    At $\mathrm{O}(q^2)$ we find \begin{equation}
D\tilde \chi = \tilde \sigma \left(1 + \partial_x \left(\frac{b_1}{\tilde\chi}\right)\right) + \partial_x \left(\tilde \sigma \partial_x \left(\frac{b_2}{\tilde \chi}\right) + \frac{ \tilde \sigma b_1}{\tilde \chi}\right). \label{eq:DappA}
\end{equation}
Again assuming $b_{1,2}$ do not diverge, we may spatially average both sides of (\ref{eq:DappA}).  Using (\ref{eq:1C0}) and (\ref{eq:1C}) along with (\ref{eq:sigmaE}) and (\ref{eq:chiE}) we recover the Einstein relation (\ref{eq:einstein}).

We conclude with a technical comment.  The true eigenstates of an inhomogeneous diffusion equation in one spatial dimension are spatially localized,  with a frequency dependent localization length that diverges as $Dq^2 \rightarrow 0$ \cite{ziman2}.   Subject to mild assumptions about the distribution of $\tilde\sigma$ \cite{ziman2},  the localization length diverges fast enough that the diffusion constant and conductivity are finite.   While the ansatz (\ref{eq:nxt}) appears to describe the time evolution of a delocalized eigenstate of (\ref{eq:Adiff}), in principle an expansion of (\ref{eq:nxt}) to all orders in $b_{0,1,2,\cdots}$ can be consistent with localization; above we only computed $b_{0,1}$ explicitly.    The assumptions which we made that $b_{1,2}$ did not diverge with $x$ amount to the assumption that the localization length grows quickly enough at low frequencies, are sensible in ordinary models \cite{ziman2}.    An alternative way to think about this problem is to assume $q\in \mathrm{e}^{\mathrm{i}\mpi /4}\mathbb{R}$, so that the driving is periodic in time and leads to a spatial decay of $n(x)$.   This spatial decay will be dominated at low driving frequencies not by the localization of eigenstates, but by dissipative diffusion \cite{lucasplasma}.

\section{Fluid-Gravity Expansion in Kruskal Coordinates}
\label{app:flugrav}
In this appendix, we carry out the fluid-gravity expansion to second order in Kruskal coordinates, for our static striped black holes.    For simplicity, we work in $d=2$ and assume that the background is governed by the Einstein-dilaton system (\ref{eq:sec5action}).   The zeroth order metric is given by (\ref{eq:kruskalBG}) and to this order, after making the change of variables \begin{equation}
A = \mathrm{e}^{\hat a}, \;\;\;\; B = \mathrm{e}^{\hat b},
\end{equation}
 we have the following equations of motion (denoting $UV\equiv \rho$, and $\partial_\rho$ with primes): 
\begin{subequations}
	\begin{align}
\mathrm{e}^{-{\hat b}}\Big(\rho \mathrm{e}^{\hat b}\phi^{\prime}\Big)^{\prime}&=\frac{1}{2}\mathrm{e}^{\hat a}\frac{\partial\mathcal{V}(\phi)}{\partial\phi}, \label{eq:eomd}
\\
\hat a^{\prime}\hat b^{\prime}-\frac{1}{2}(\hat b^{\prime})^{2}-\frac{1}{2}(\phi^{\prime})^{2}-\hat b^{\prime\prime}&=0, \label{eq:eom1}
\\
\frac{1}{2}\mathrm{e}^{\hat a}\mathcal{V}(\phi)+\rho \hat b^{\prime 2}+\hat b^{\prime}+\rho \hat b^{\prime\prime}&=0,\label{eq:eom2}
\\
\frac{1}{2}\mathrm{e}^{\hat a}\mathcal{V}(\phi)+\frac{1}{2}\rho \phi^{\prime 2}+\frac{1}{2}\rho \hat b^{\prime 2}+\hat a^{\prime}+\hat b^{\prime}+\rho \hat a^{\prime\prime}+\rho \hat b^{\prime\prime}&=0. \label{eq:eom3}
\end{align}
\end{subequations}
As we have explained in the main text, subleading corrections to these equations in $\xi^{-1}$ are regular everywhere, and hence this forms the basis for a well-behaved perturbative expansion.   As we go through this appendix, we will see explicitly what these regular corrections are.

\subsection{First Order Correction}
By parity symmetry under a local change $x\rightarrow -x$, and due to the fact that there is a Killing vector $V\partial_U - U\partial_V$,   the only corrections to the metric which could arise at $\mathrm{O}(\xi^{-1})$ are
\begin{equation}\label{eq:kruskalFG}
\mathrm{d}s^{2}_{1} =  \zeta(UV)(V\mathrm{d}U\mathrm{d}x+ U\mathrm{d}V\mathrm{d}x)
\end{equation}
These are simply coordinate artifacts.    To see this, we note that at first order, \begin{align}
0 &= \frac{1}{V} \left(R_{Ux} - \frac{R}{2}g_{Ux} - 8\mpi G_{\mathrm{N}}T_{Ux}\right) \notag \\
&= \zeta \mathrm{e}^{-\hat a} \left[\frac{\mathrm{e}^{\hat a}\mathcal{V}}{2} + \hat a^\prime + \hat b^\prime + \rho\left(\hat a^{\prime\prime}+\hat b^{\prime\prime}\right) + \frac{\rho}{2}\left(\hat b^{\prime2} + \phi^{\prime 2}\right)\right] + \frac{\hat b^\prime \partial_x \hat a - \phi^\prime \partial_x \phi - \partial_x(\hat a^\prime + \hat b^\prime)}{2}  .  \label{eq:ein1st}
\end{align}
Using (\ref{eq:eom3}) the coefficient of $\zeta$ identically vanishes.  Indeed, this is a consequence of the fact that $\zeta$ can be removed by a change in coordinates, so for simplicity we set it to vanish.   There is another contribution which is $\zeta$-independent.   Let us now consider the following particular radial derivative: \begin{align}
\mathrm{e}^{-\hat b}&\left[\rho \mathrm{e}^{\hat b}\left(\hat b^\prime \partial_x \hat a - \phi^\prime \partial_x \phi - \rho \partial_x(\hat a^\prime + \hat b^\prime)\right)\right]^\prime = - \partial_x \phi \left[ \phi^{\prime\prime} + (1+\rho \hat b^\prime)\phi^\prime\right] - \partial_x \left(\rho \frac{\phi^{\prime 2}}{2} + \rho \hat a^{\prime\prime} + \rho \hat b^{\prime\prime} + \hat a^\prime + \hat b^\prime + \frac{\rho}{2}\hat b^{\prime 2} \right) \notag \\
& + \partial_x \hat a \left(\hat b^\prime + \rho \hat b^{\prime 2} + \rho \hat b^{\prime\prime}\right) = -\frac{\mathrm{e}^{\hat a}}{2} \frac{\partial \mathcal{V}}{\partial \phi} \partial_x \phi + \partial_x \left(\frac{\mathrm{e}^{\hat a}}{2} \mathcal{V}\right) - \frac{\mathrm{e}^{\hat a}}{2} \mathcal{V}\partial_x a = 0.
\end{align}
In the last step above we have used the zeroth order equations of motion (\ref{eq:eomd}), (\ref{eq:eom3}) and (\ref{eq:eom2}) respectively.    We see that (up to a factor $\rho \mathrm{e}^{\hat b}$) the non-vanishing contribution to (\ref{eq:ein1st}) is constant in $\rho$.  We may evaluate it at the horizon, $\rho=0$ -- since all fields are regular at the horizon on the fluid-gravity ansatz, we conclude that the entire contribution in square brackets above vanishes.   Hence, away from the horizon, where $\rho \mathrm{e}^{\hat b}$ is strictly finite, we conclude that the remaining contribution to (\ref{eq:ein1st}) vanishes.   Thus, there is no first order correction to our geometry.

\subsection{Second Order Correction}
The first nonvanishing corrections to the metric and dilaton fields occur at second order and are of the following form 
\label{app:kruskal}
\begin{subequations}
	\begin{align}
	\mathrm{d}s^{2}_2 &=V^{2}\eta(UV)\mathrm{d}U^{2}+ U^{2}\eta(UV)\mathrm{d}V^{2}+\alpha(UV)\mathrm{d}U \mathrm{d}V+[\beta(UV) + \gamma(UV)] \mathrm{d}x^{2} \notag \\
	&\;\;\;\;\;\;\; +[\beta(UV)-\gamma(UV)] \mathrm{d}y^{2}+ \mathrm{O}\left(\frac{1}{\xi^{3}}\right) \label{eq:kruskalFG}
	\\
	\Phi_2 &=\phi_2(UV)
	\end{align}
\end{subequations}
After using the background equations of motion to simplify the results somewhat, we obtain the following equations of motion for the perturbations:\begin{subequations}\label{eq:89}
	\begin{align}
	\gamma &\left[\frac{B^\prime}{B} - \frac{\rho B^{\prime 2}}{B^2} + \frac{\rho B^{\prime\prime}}{B}\right] + \gamma^\prime\left(\frac{\rho B^\prime}{B}-1\right) -\gamma^{\prime\prime} = -\frac{(\partial_x A)^2}{4A} - \frac{\partial_x A \partial_x B}{2B} - \frac{A}{4}(\partial_x \phi)^2 - \frac{\partial_x^2 A}{2} \\
	\frac{2B}{A^2}\eta &\left(\frac{\rho A^\prime}{A}-1\right) + \left(\frac{\rho^2 B A^\prime}{A^3} - \frac{4\rho B}{A^2} - \frac{\rho^2 B^\prime}{A^2}\right)\eta^\prime - \frac{\rho B^{\prime\prime}}{A^2}\eta  - \beta \left[\frac{\rho B^{\prime\prime} + B^\prime}{AB} - \frac{\rho B^{\prime 2}}{AB^2}\right] \notag \\
	&+\frac{\beta^\prime}{A}\left(1-\frac{\rho B^\prime}{B}\right) + \frac{\rho}{A}\beta^{\prime\prime}  - \alpha \left[ \frac{\rho B^{\prime\prime}}{A^2} + \frac{\rho B^{\prime 2}}{2A^2B} + \frac{B^\prime}{A^2} - \frac{\rho BA^{\prime 2}}{A^4}\right]  + \frac{B\alpha^\prime}{A^2}\left(1-\frac{2\rho A^\prime}{A}\right)+ \frac{\rho B}{A^2} \alpha^{\prime\prime} \notag \\
	&+ \frac{B}{2} \frac{\partial \mathcal{V}}{\partial \phi} \phi_2 + \frac{\rho B \phi^\prime \phi_2^\prime}{A} = - \frac{\partial_x^2 A}{2A}. \\
	\beta &\left[\frac{B^{\prime\prime}}{B^2} - \frac{A^\prime B^\prime}{AB^2}  - \frac{B^{\prime 2}}{B^3} \right] + \left(\frac{A^\prime }{AB} + \frac{B^\prime}{B^2}\right)\beta^\prime - \frac{\beta^{\prime\prime}}{B} - \alpha \frac{A^\prime B^\prime}{A^2B} + \frac{B^\prime \alpha^\prime}{AB} - \phi^\prime \phi_2^\prime = 0 \\
	\frac{B\mathcal{V}\alpha}{2} &- \frac{\beta}{B}\left(B^\prime + \rho B^{\prime\prime}\right) + \beta^\prime + \rho \beta^{\prime\prime} - \left(\frac{\rho^2\eta B^\prime}{A}\right)^\prime + \frac{BA}{2} \frac{\partial \mathcal{V}}{\partial \phi}\phi_2^\prime = \frac{(\partial_x A)^2}{4A} + \frac{A(\partial_x B)^2}{2B^2} - \frac{A(\partial_x \phi)^2}{4} \notag \\
	&-\frac{\partial_x^2 A}{2} - \frac{A\partial_x^2 B}{2B} \\
	\frac{2\rho^2 A^\prime \phi^\prime}{A^3}\eta &- \frac{2\rho^2 \phi^\prime \eta^\prime}{A^2} -\frac{2\rho B^\prime \phi^\prime}{AB^2}\beta + \frac{2\rho \phi^\prime}{AB}\beta^\prime +\frac{2\rho}{A}\phi_2^{\prime\prime} + \frac{2\rho B^\prime \phi_2^\prime}{AB} + \frac{2\phi_2^\prime}{A} - \frac{\partial^2 \mathcal{V}}{\partial \phi^2}\phi_2 = -\frac{\partial_x^2 \phi}{B} - \frac{\partial_x A\partial_x\phi}{AB}
	\end{align}
\end{subequations}
As in the standard fluid-gravity correspondence, we see that these perturbations obey ordinary differential equations depending on $\rho$ alone, pointwise at each $x$.   

Although we are not able to solve these equations analytically, we do note that they are completely regular as $\rho \rightarrow 0$ (near the black hole horizon), supporting our claim that this expansion of Einstein's equations is well-behaved.   We also note that there is some gauge redundancy in the above equations of motion.   One useful gauge will be to set $\eta=0$ -- this can be done through a coordinate change of the form $U\rightarrow U[1+\Xi (UV,H(x))]$, $V\rightarrow V[1+\Xi(UV,H(x))]$ for a small $\Xi(UV,H(x)) \sim \xi^{-2} $.

\section{Higher Order Corrections to the Butterfly Velocity}
\label{app:vbcorrections}
In this appendix, we will assume that we have found the regular solution to the fluid-gravity expansion in Kruskal coordinates up to second order.   We choose the gauge $\eta=0$, which simplifies the computation of the shock wave geometry.

We can now use (\ref{eq:shock}) to calculate the butterfly velocity to next order for $d=2$.   The only additional metric correction at $\mathrm{O}\left(\xi^{-2}\right)$ comes in the $UU$ component of Einstein's equation.  After employing (\ref{eq:UDelta}),  only the $UU$ component of Einstein's equation is altered by the shockwave. Looking at this component, and demanding that the coefficient of $\mdelta(U)$ vanishes, we find the differential equation \begin{equation}
\left(1+\frac{\alpha}{A}-\frac{\beta+\gamma}{B}\right)\partial_x^2 h + \frac{\partial_x A\partial_x h}{A} - \left(\frac{B^\prime}{A} + \frac{B}{A}\left(\frac{\beta}{B}\right)^\prime \right)h = 8\mpi G_{\mathrm{N}}\frac{B(0,0)}{A(0,0)} \mathcal{E}\mathrm{e}^{2\mpi Tt} \mdelta(x) \label{eq:heqappc}
\end{equation}
after employing (\ref{eq:89}), at $\rho=0$, to simplify the result.     Following the derivation in the main text, this leads to an effective butterfly velocity \begin{equation}
\frac{1}{v_{\textsc{b}}} = \frac{1}{2\mpi T} \mathbb{E}\left[\tilde m\right],
\end{equation}where \begin{equation}
\tilde m^2 = \frac{B^\prime}{A} \left[1-\frac{\alpha}{A} + \frac{\beta+\gamma}{B}\right] + \frac{B}{A}\left(\frac{\beta}{B}\right)^\prime + \frac{1}{\sqrt{A}}\partial_x^2 \sqrt{A}+ \mathrm{O}\left(\frac{1}{\xi^3}\right). 
\end{equation}
To get this result, we have divided through (\ref{eq:heqappc}) by the coefficient of $\partial_x^2 h$, and then used that (for $x\ne 0$) \begin{align}
&\partial_x^2 h + \frac{\partial_x A\partial_x h}{A} -  \left[\frac{B^\prime}{A} \left[1-\frac{\alpha}{A} + \frac{\beta+\gamma}{B}\right] + \frac{B}{A}\left(\frac{\beta}{B}\right)^\prime\right] h \notag \\
&= \partial_x^2 \left(h\sqrt{A}\right) - h\sqrt{A}\left\lbrace\left[\frac{B^\prime}{A} \left[1-\frac{\alpha}{A} + \frac{\beta+\gamma}{B}\right] + \frac{B}{A}\left(\frac{\beta}{B}\right)^\prime\right] + \frac{1}{\sqrt{A}}\partial_x^2 \sqrt{A}\right\rbrace =   \mathrm{O}\left(\frac{1}{\xi^3}\right).
\end{align}

  \end{appendix}

\addcontentsline{toc}{section}{References}
\bibliographystyle{unsrt}
\bibliography{butterflybibcopy}
\end{document}